\documentclass[letterpaper, 10pt, conference]{ieeeconf}  
\IEEEoverridecommandlockouts
\usepackage[top=0.75in, bottom=1in, left=0.75in, right=0.75in]{geometry} 
\usepackage{cite}
\usepackage{url}
\usepackage{amsmath,amssymb,amsfonts}
\usepackage{gensymb}
\usepackage{siunitx}
\usepackage{graphicx}
\usepackage{booktabs}
\usepackage{textcomp}
\usepackage{mathtools}
\usepackage{graphicx}
\usepackage[caption=false]{subfig}
\usepackage{xcolor}
\usepackage[nolist,nohyperlinks]{acronym}
\def\BibTeX{{\rm B\kern-.05em{\sc i\kern-.025em b}\kern-.08em
    T\kern-.1667em\lower.7ex\hbox{E}\kern-.125emX}}

\begin{document}

\title{\LARGE \bf
A Non-linear Differentiable Model for\\Stormwater-based Irrigation of a Green Roof in Toronto
}

\author{Chia-Hui Yeh and Margaret P. Chapman, \emph{Member}, \emph{IEEE}
\thanks{C.H.Y. and M.P.C. are with the Edward S. Rogers Sr. Department of Electrical and Computer Engineering, University of Toronto, 10 King's College Road, Toronto, Ontario M5S 3G8, Canada. Contact emails: {\tt\small chiahui.yeh@mail.utoronto.ca, mchapman@ece.utoronto.ca}.}
\thanks{C.H.Y. was supported by a Natural Sciences and Engineering Research Council of Canada Undergraduate Student Research Award. C.H.Y. and M.P.C. also gratefully acknowledge support from the University of Toronto.}
}

\maketitle

\begin{abstract}
Green infrastructure has potential to alleviate the environmental impact of rapidly growing cities. This potential has inspired laws in Toronto that require the inclusion of rooftops with large vegetation beds, called \emph{green roofs}, into sufficiently sized construction projects. We study the problem of reusing stormwater to irrigate a green roof in Toronto, where potable water is the current irrigation source. The vision is that widespread reuse of stormwater runoff for irrigation of green roofs and other purposes can reduce sewer overflow volumes without over-building (with the added benefit of conserving potable water). Towards this vision, our goal is to develop and evaluate two pump controllers for transporting stormwater to the green roof of interest in simulation. A key contribution is our development of a site-specific non-linear model for stormwater flow using smoothing techniques that permits linearization and a standard model predictive controller (MPC). We compare the efficacy of the MPC, which anticipates the weather, and an on/off controller, which is reactive rather than anticipative, for the site in simulation. With further study, we are hopeful that this research will advance control systems technology to improve the performance of green and stormwater infrastructure in growing urban areas. 
\end{abstract}

\section{Premise}
The incorporation of green infrastructure into urban centers has been reported to provide benefits to the environment, economy, and human welfare \cite[Tables 1--4]{parker2019green}. 
%
For example, the installation of green roofs has potential to reduce the maximum discharge rate and the total water volume released by a drainage network during periods of rainfall \cite{ERCOLANI2018830}.
Different designs for green roofs have different advantages and disadvantages regarding their ability to retain stormwater and improve its quality \cite{BEECHAM2015370}.  
Motivated by urban drainage benefits, Raimondi and Becciu have developed a probabilistic approach to assess the performance of green roof systems, in which the system may contain stormwater initially \cite{raimondi_becciu_2020}. 

Growing evidence for the benefits of green infrastructure has motivated some city governments to pass laws that require new or retrofitted construction to include a green infrastructure component. For example, the City of Toronto passed a regulation in 2009, specifying the inclusion of green roofs into new or modified infrastructure that covers a large enough area \cite{greenroofbylaw}. The regulation calls for ``adequate measures...to permit irrigation necessary to initiate and sustain the vegetation during the service life of the green roof'' \cite[Article IV, Sec. 492-9 M]{bylawdoc}. This regulation motivates investigations about which measures may be more appropriate to sustain the vegetation, and here, we undertake an investigation from the perspective of automatic control.

Passive control has the advantage of reduced financial and maintenance requirements compared to automatic control. 
However, the role of automatic control in stormwater management and other water resources applications is growing. 
Romero et al. has surveyed automatic control systems for agricultural applications up to 2012 and has compared model predictive controllers and proportional-integral-derivative controllers in simulation \cite{romero2012research}. More recently, in 2021, a real-time irrigation method that detects soil moisture has improved the efficiency of water usage for cultivating tomatoes in a greenhouse
\cite{liao2021development}.
In prior work, we have used a safety analysis method based on robust optimal control to assess different designs for stormwater systems numerically \cite{chapman2018reachability}. Model predictive control has been applied in simulation to alleviate the severity of coastal flooding in Norfolk, Virginia using the United States Environmental Protection Agency Storm Water Management Model (SWMM) \cite{sadler2020exploring}. 

%

Moreover, researchers have investigated methods for automatic sensing and control of green roof systems. From 2002 to 2006, the performance of green roofs in colder climates was examined by the Toronto and Region Conservation Authority, and the study used automated devices to sample runoff from different roof surfaces for assessing water quality \cite[p. 5]{yorkreport}. 
With a focus on the eastern Texas climate, Aydin et al. has envisioned a green roof system that gathers daily weather data to develop watering schedules and uses wastewater for irrigation \cite[Sec. 4]{Aydin2018}. 
In research about the seasonal aspects of green roof performance, an automatic sprinkler system has provided irrigation until a sensor detects a sufficient rainfall depth \cite{peng2015seasonal}. By the same research group, a neural network has been applied to relate weather data to soil moisture to estimate a green roof's water demand and operate an on/off irrigation controller \cite{TSANG2016360}. The approach has been validated experimentally at a site in Hong Kong \cite{TSANG2016360}. 
Moreover, an automated pump can facilitate the reuse of stormwater runoff for irrigating a green roof, which applies to our case study of interest. 

\begin{figure}[ht]
\centerline{\includegraphics[width=\columnwidth]{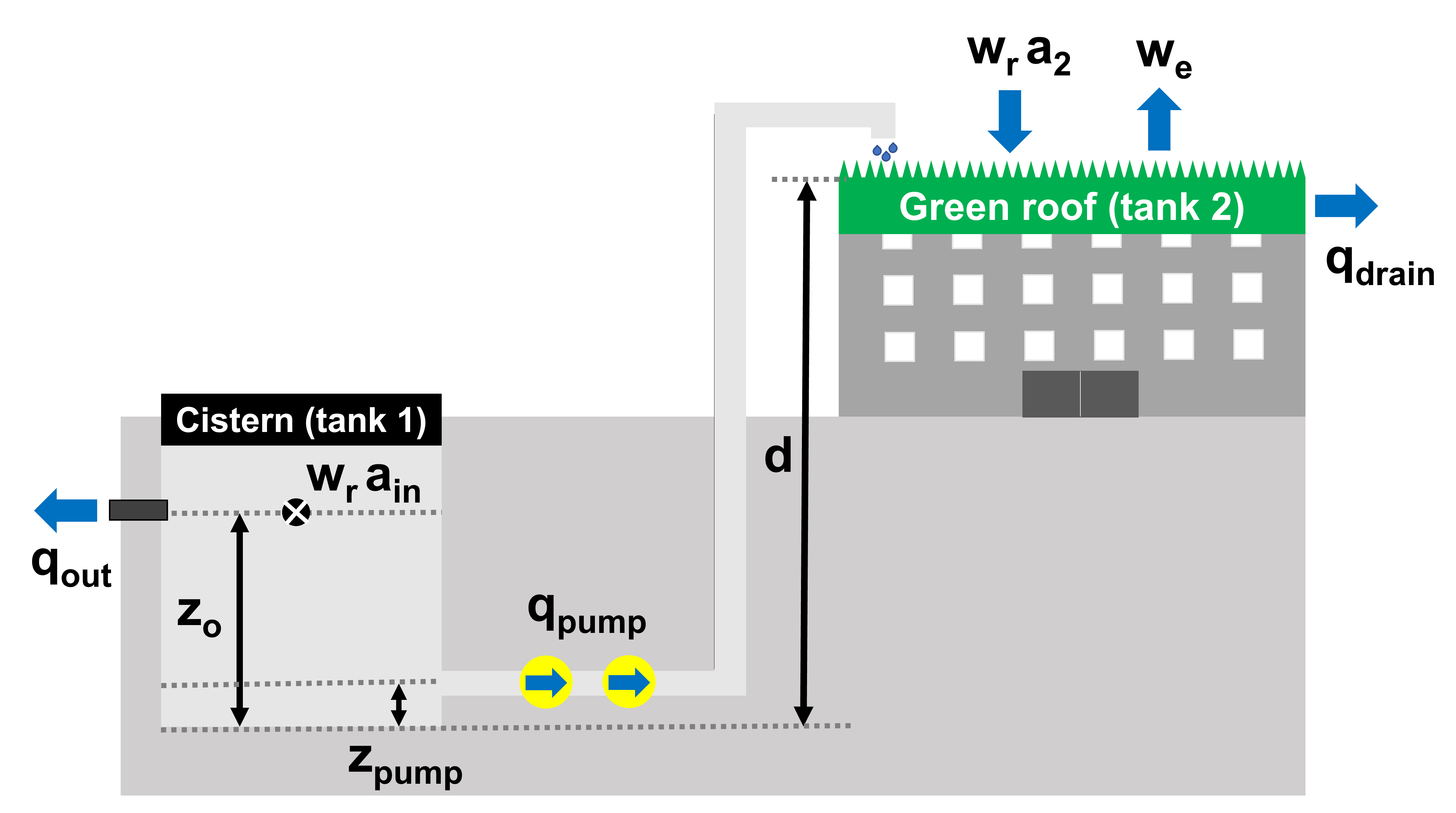}}
\caption{A schematic of a green roof system in Toronto (Spadina Ave., GRIT Lab, University of Toronto). This system consists of an underground cistern that collects stormwater, a green roof with vegetation, and two pumps in series for transporting stormwater to irrigate the vegetation. In the real system, potable water is the current irrigation source.}
\label{dynamic-system-fig}
\end{figure}

\textit{Research aims.} We consider a system consisting of an underground cistern that collects stormwater runoff, a green roof that requires irrigation, and two pumps in series (Fig. \ref{dynamic-system-fig}). The system is part of the University of Toronto's Green Roof Innovation Testing Laboratory (GRIT Lab), a facility established in 2010 that studies the performance of green roofs, green walls, and photovoltaic arrays. The pumps have been installed with the intention of reusing stormwater for irrigation; however, potable water is the current irrigation source. Like other cities in North America, Toronto's sewers can release untreated wastewater into natural waterways during heavy storms \cite{torontocso}. Widespread stormwater reuse (e.g., for irrigation, toilet flushing, cooling) could increase the effective capacity of Toronto's sewers and thereby reduce overflow volumes without over-building. Towards this long-term aim, our short-term aim is to develop a model predictive controller that anticipates the weather and assess its efficacy relative to an on/off controller for stormwater-based irrigation of the green roof at hand in simulation. 

\textit{Contributions.} We report two contributions. Our first contribution is to devise a \emph{site-specific non-linear differentiable model} for stormwater flow through the green roof system. We use site-specific engineering drawings, including a pump performance curve; we employ sigmoids to estimate case statements in a differentiable manner; and we adopt a differentiable approximation for the square root \cite{duan2016smoothing}.
We use our green roof system model, data from a Toronto weather station, and an evapotranspiration model \cite{zotarelli2010step} to develop a model predictive controller (MPC). In parallel, we propose an on/off controller that reacts to the current water levels in the cistern and green roof. The MPC is parametrized by a weight that penalizes high pumping rates; the on/off controller is parametrized by a multiple of the maximum pumping rate. We assess the performance of each controller for various parameter settings and initial states by simulating a 12-hour period of wet weather from eastern Canada. Our second contribution is to demonstrate that with proper parameter settings, the MPC outperforms the on/off controller. Given that the MPC can adjust the pumping rate continuously while considering a forecast, the water depth in the green roof can be maintained closer to a desired level.

\section{Process}
First, we present the development of the model predictive controller and then we present the on/off controller.

\subsection{Model Predictive Controller}
We derive a non-linear model for the flow of stormwater through the green roof system using simplified Newtonian physics. We employ this model to design a model predictive controller that uses weather data to optimize the pumping rate. An interesting feature of our approach is the application of sigmoid functions and a smooth square-root approximation to enhance the model's analytical properties.
Here, we present a non-linear non-differentiable model, a differentiable continuous approximation, a linearization procedure, and a model predictive control algorithm. 

\subsubsection{Non-linear non-differentiable model}\label{dynamic-system-section}
The physical system of interest is the flow of stormwater from the cistern to the green roof, where the flow is controlled by two pumps in series (Fig. \ref{dynamic-system-fig}). 
To form a dynamical model, we represent the cistern and the green roof as two tanks, tank 1 and tank 2, respectively.
We construct our model using a mass balance of water entering or leaving tank 1 or tank 2.
For tank 1, the source of water inflow is the precipitation $w_\text{r}$ (m/s) through an inlet of area $a_\text{in}$ (m\textsuperscript{2}), i.e., surface runoff from the street. 
Water leaves tank 1 when the height of water exceeds $z_\text{o}$ (m), and $q_\text{out}$ (m\textsuperscript{3}/s) is the volumetric discharge rate through the outlet of tank 1 (to a sewer downstream that is not modeled explicitly). 
For tank 2, the sources of water inflow include the precipitation $w_\text{r}$ and the aggregate irrigation from the two pumps in series $q_\text{pump}$ (m\textsuperscript{3}/s);
the sources of water outflow include evapotranspiration $w_\text{e}$ (m\textsuperscript{3}/s) and drainage from the vegetation due to the soil capacity $q_\text{drain}$ (m\textsuperscript{3}/s).
Table \ref{table-params} lists model parameters. 

\begin{figure}[ht]
\centerline{\includegraphics[width=\columnwidth]{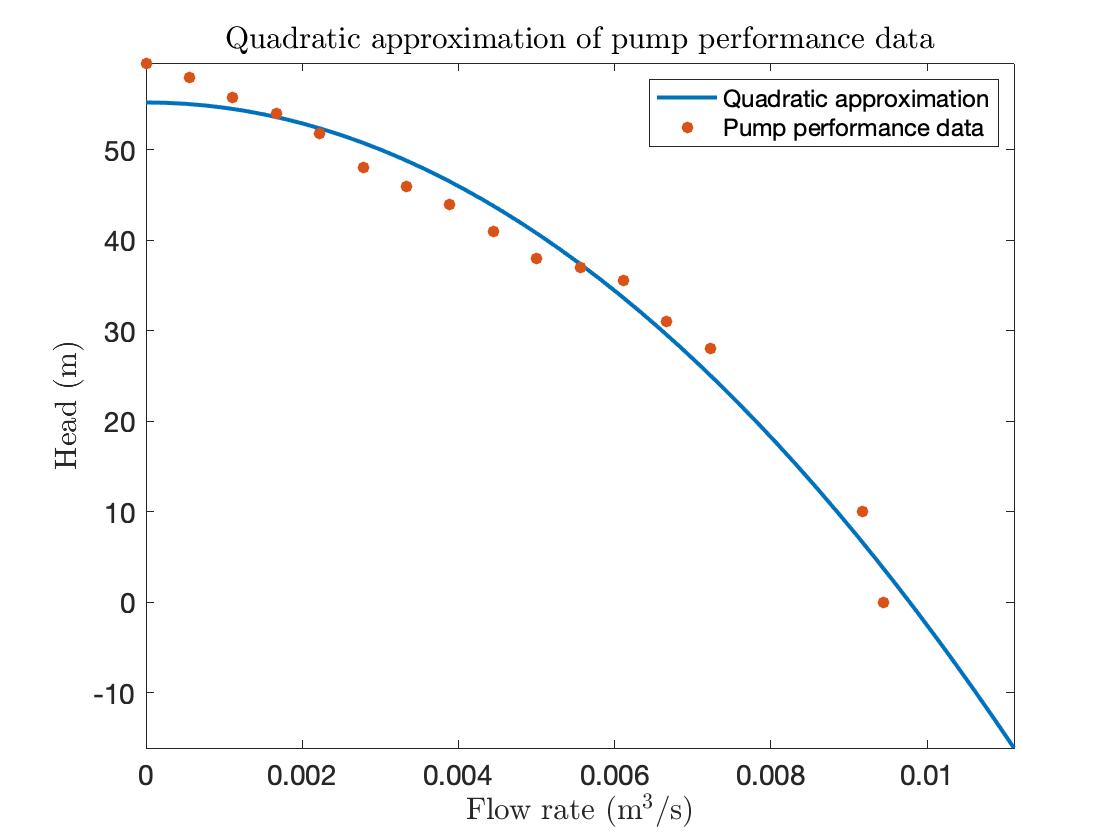}}
\caption{The figure shows measurements from visually inspecting a given pump performance curve (circles) and our quadratic fit $\phi(y) \coloneqq \hat a y^2 + \hat c$ \eqref{quadapprox} to the measurements (solid blue).}
\label{pump-curve}
\end{figure}

Our model for the system consists of three key entities: the state, control, and disturbance. The \emph{state} at time $t$ is a vector $x_t \coloneqq [x_{t,1}, x_{t,2}]^T \in \mathbb{R}^2$, where $x_{t,i}$ (m\textsuperscript{3}) is a volume of water in tank $i$. The \emph{control} at time $t$ is a proportion of the maximum aggregate flow rate produced by the pumps $u_t \in \mathbb{R}$ (no units). The \emph{disturbance} at time $t$ is a vector $w_t \coloneqq [w_{t,\text{r}},w_{t,\text{e}}]^T \in \mathbb{R}^2$, where $w_{t,\text{r}}$ (m/s) is a precipitation rate and $w_{t,\text{e}}$ (m\textsuperscript{3}/s) is a volumetric evapotranspiration rate of the vegetation due to solar irradiance and other climate factors \cite{zotarelli2010step}.
%
%
We assume that the disturbance is known exactly and the current state is fully observable for simplicity in this work. The model is given by
%
\begin{subequations}\label{fullequationmodel}
\begin{equation}\begin{aligned}
    x_{t+1} = x_t + \tau f(x_t,u_t,w_t), \;\;\; t = 0,1,2, \dots,
\end{aligned}\end{equation}
where $\tau$ is the duration of an interval $[t, t+1)$,
%
%
$f$ is chosen according to simplified Newtonian physics,
\begin{equation}\label{system}\begin{aligned}
f(x,u,w) & \coloneqq \left[ f_{1}(x,u,w), f_{2}(x,u,w) \right]^T\\
f_{1}(x,u,w) & \coloneqq  w_{\text{r}}\cdot a_{\text{in}} - q_{\text{out}}(x_{1}) - q_\text{pump}(x,u)\\
f_{2}(x,u,w) & \coloneqq  w_{\text{r}}\cdot a_{2} + q_\text{pump}(x,u)  - w_\text{e} - q_\text{drain}(x_2),
\end{aligned}\end{equation}
\end{subequations}
$x \coloneqq [x_1,x_2]^T \in \mathbb{R}^2$, $w \coloneqq [w_{\text{r}},w_{\text{e}}]^T \in \mathbb{R}^2$, and we define $q_\text{out}$, $q_\text{pump}$, and $q_\text{drain}$ subsequently.
%
The cistern is equipped with a gravity-driven outlet with elevation $z_\text{o}$ (m) and radius $r_\text{o}$ (m). The discharge rate (m\textsuperscript{3}/s) through this outlet is given by
\begin{equation}\label{qout}
\begin{aligned}
q_\text{out}(x_{1}) \hspace{-.5mm} \coloneqq \hspace{-.5mm}  \begin{cases} c_{\text{out}} \sqrt{ \frac{x_{1}}{a_{1}} - z_\text{o}} & \text{if } \frac{x_{1}}{a_{1}} > z_\text{o} \\0 & \text{otherwise} \end{cases}, \;\; c_{\text{out}}  \hspace{-.5mm} \coloneqq \hspace{-.5mm}  c_\text{d}  \pi r_\text{o}^2  \sqrt{2 g}.
\end{aligned}
\end{equation}
The aggregate flow rate generated by the pumps $q_{\text{pump}}(x,u)$ (m\textsuperscript{3}/s) is a proportion of the maximum flow rate $\bar{q}_{\text{pump},0}(x)$ (m\textsuperscript{3}/s). The expression for $q_{\text{pump}}(x,u)$ is given by
\begin{subequations}\label{qpump}
\begin{equation}\label{physics-pump}\begin{aligned}
q_{\text{pump}}(x,u) & \coloneqq \begin{cases} 0 & \text{if }
\frac{x_{2}}{a_{2}} \geq z_\text{veg} \text{ or} \\ 
\text{} & \frac{x_{1}}{a_{1}} < z_\text{pump}+ z_\text{H}\\
u \cdot \bar{q}_{\text{pump},0}(x) & \text{otherwise}, \end{cases}
\end{aligned}\end{equation}
where $z_\text{H}$ (m) is the minimum water level relative to a pump's base that pumping requires (called \emph{net positive suction head}), and 
$z_\text{veg}$ (m) is the desired water depth to ensure sufficient soil moisture for the vegetation. That is, if the vegetation does not require irrigation ($\frac{x_{2}}{a_{2}} \geq z_\text{veg}$) or if there is not enough water in the cistern for pumping
($\frac{x_{1}}{a_{1}} < z_\text{pump} + z_\text{H}$), then the pumps produce zero flow rate. Otherwise, the flow rate is proportional to the maximum flow rate, where the proportion is given by the control $u$.

We have derived the maximum flow rate $\bar{q}_{\text{pump},0}(x)$ using a quadratic approximation for a pump's performance curve (Fig. \ref{pump-curve}) and the total head loss of the pipe that is connected to the pumps. We have used the quadratic approximation
\begin{equation}\label{quadapprox}
    \phi(y) \coloneqq \hat a y^2 + \hat c,
\end{equation}
where $\phi(y)$ (m) is the head and $y$ (m\textsuperscript{3}/s) is the flow rate produced by a pump. We have measured the head and flow rate by inspecting a given performance curve, and we have multiplied the flow rate measurements by two because the site has two pumps in series. Then, we have fitted the coefficients $(\hat a, \hat c)$ \eqref{quadapprox} to our measurements via least-squares minimization (Fig. \ref{pump-curve}). A key feature of \eqref{quadapprox} is the absence of a term that is linear in $y$. This modeling choice simplifies the derivation of the maximum flow rate because the total head loss also lacks a term that is linear in the flow rate.
%

The \emph{total head loss} (m) is a sum of the head loss terms,
\begin{equation}\label{totalheadloss}
    L(x,y) \coloneqq  \underbrace{ \textstyle (F \frac{l}{D} + k_L ) \cdot \frac{y^2}{2g a_{\text{pump}}^2}}_{\text{friction and minor head losses}} + \underbrace{\textstyle d - \frac{x_1}{a_1}.}_{\text{approx. static head loss}}
\end{equation}
 We have evaluated the \emph{friction and minor head losses} using the geometry of the pipe and the expression given by \cite[Eq. 6.79, p. 389]{white2011fluid}. The \emph{static head loss} is the total vertical distance that the water is raised by the pump, which equals $d - \frac{x_1}{a_1} + \frac{x_2}{a_2}$. In \eqref{totalheadloss}, we have neglected the term $\frac{x_2}{a_2}$ for simplicity because it is substantially smaller than $d$; $\frac{x_2}{a_2}$ is the level of water in the green roof, whereas $d$ is larger than the height of the building (Fig. \ref{dynamic-system-fig}). Suppose that $\frac{x_1}{a_1} \geq z_{\text{pump}} + z_{\text{H}}$ holds. Then, we model the maximum flow rate $\bar{q}_{\text{pump},0}(x)$ as the non-negative solution $y$ to the quadratic equation, 
 \begin{equation}\label{intersection}
     \phi(y) = L(x,y),
 \end{equation}
where $\phi$ and $L$ are given by \eqref{quadapprox} and \eqref{totalheadloss}, respectively.
Equation \eqref{intersection} represents how the operation of the pumps is affected by their placement underground. By rearranging the terms in \eqref{intersection}, the maximum flow rate $\bar{q}_{\text{pump},0}(x)$ is given by
%
%
%
\begin{equation}\label{physics-pump-sub}
\bar{q}_{\text{pump},0}(x) \coloneqq b \sqrt{\textstyle \frac{x_1}{a_1} + \hat c - d}, \;\;\; b \coloneqq \left(\textstyle \frac{(F \frac{l}{D} + k_\text{L})}{2 g a_{\text{pump}}^2}-\hat a\right)^{-1/2}.
\end{equation}
\end{subequations}
Note that $\frac{x_1}{a_1} \geq z_{\text{pump}} + z_{\text{H}}$ implies that $\bar{q}_{\text{pump},0}(x) > 0$ holds, which we have verified using the values in Table \ref{table-params}.

For tank 2, water in the soil drains when the soil reaches its capacity limit. This phenomenon is modeled with $q_\text{drain}$ (m\textsuperscript{3}/s) using Darcy's Law \cite[Example 2-2, Case A, pp. 142--144]{selker_or_2018} as follows:
\begin{equation}\label{qdrain}\begin{aligned}
    q_\text{drain}(x_2) \coloneqq \begin{cases} 0 & \text{if }
    x_2 < z_\text{cap} \\
    K \cdot a_2 \cdot \frac{(\frac{x_2}{a_2} + z_\text{soil})}{z_\text{soil}} & \text{otherwise,} \end{cases}\\
\end{aligned}\end{equation}
where $K$ (m/s) is the saturated hydraulic conductivity, $z_\text{soil}$ (m) is the depth of the soil on the green roof, and $z_\text{cap}$ (m\textsuperscript{3}) is the volumetric capacity of the soil.

\begin{table}[ht]
\caption{Green roof system parameters}
\label{table-params}
\setlength{\tabcolsep}{3pt}
\begin{tabular}{|p{25pt}|p{125pt}|p{60pt}|}
\hline
\textbf{Symbol} &  \textbf{Description} & \textbf{Value} \vspace{.5mm}\\
\hline
$a_1$ & Bottom surface area of the cistern & 25 m\textsuperscript{2} \vspace{.6mm}\\
$a_2$ & Bottom surface area of the rooftop vegetation & 68.8 m\textsuperscript{2} \vspace{.6mm}\\
$a_\text{pump}$ & Area of flow through the pump & $0.01 \cdot \pi$ m\textsuperscript{2} \vspace{.6mm}\\ 
$a_\text{in}$ & Area of flow through tank 1's inlet & (0.305)\textsuperscript{2}$\cdot\pi$ m\textsuperscript{2} \vspace{.6mm}\\ 
$\hat a$ & Coefficient for \eqref{quadapprox} & $-5.78 \cdot 10^{5}$ s\textsuperscript{2}$/$m\textsuperscript{5}\\
$\hat c$ & Coefficient for \eqref{quadapprox} & 55.2 m \vspace{.6mm}\\
$c_\text{d}$ & Discharge coefficient & 0.61 (no units) \vspace{.6mm}\\
$d$ & Elevation of the rooftop vegetation relative to the cistern & 16 m \vspace{.6mm}\\
$D$ & Diameter of the pipe & 0.2 m \vspace{.6mm}\\
$\epsilon$ & Positive number that dictates the steepness of a sigmoid function & 0.5 (no units) \vspace{.6mm}\\
$F$ & Friction factor of the pipe & 3.56 (no units) \vspace{.6mm}\\ 
$g$ & Acceleration due to gravity & 9.81 m/s\textsuperscript{2} \vspace{.6mm}\\ 
$K$ & Saturated hydraulic conductivity & 7.83$\cdot 10^{-8}$ \hspace{-1mm} m/s \cite[Table 1, Sandy silt, slight clay soil]{al-kharabsheh_azzam_2019} \vspace{.6mm} \\ 
$k_\text{L}$ & Minor loss coefficient of the pipe & 0.6 (no units) \vspace{.6mm}\\
$l$ & Length of the pipe & 18.4 m \vspace{.6mm}\\
$M$ & Length of the look-ahead time horizon for the model predictive controller & 10 time points \textcolor{white}{sp} ($=$ 10 s) \vspace{.6mm}\\
$N$ & Length of the time horizon during which the green roof system operates & 43200 time points ($=$ 12 h) \vspace{.6mm}\\
$\pi$ & Circle circumference-to-diameter ratio & $\approx$ 3.14 \vspace{.6mm}\\
$r_\text{o}$ & Radius of the outlet of the cistern & $0.125$ m \vspace{.6mm}\\
$ \tau $ & Duration of $[t,t+1)$ & 1 s \vspace{.6mm}\\
$x_2^*$ & Desired water volume in the green roof & $a_2 \cdot z_\text{veg}$ $\text{m}^3$ \vspace{.6mm}\\
$z_\text{cap}$ & Soil capacity &  $a_2 \cdot z_\text{soil}$ $\text{m}^3$ \vspace{.6mm}\\
$z_\text{H}$ & Minimum head that is needed for the pumps to operate & 0.6 m 
\\
$z_\text{o}$ & Elevation of the outlet of the cistern & 3 m \vspace{.6mm}\\ 
$z_\text{pump}$ & Pumps' elevation w.r.t. the base of the cistern & 0.15 m \vspace{.6mm}\\
$z_\text{soil}$ & Soil depth of the green roof & 0.5 m \vspace{.6mm} \\
$z_\text{veg}$ & Desired water depth to ensure sufficient soil moisture & $4.57 \cdot 10^{-2}$ m \cite[Table 2, Ex. I]{melvin_yonts_2009} \vspace{.6mm} \\ 
%
\hline
\multicolumn{3}{p{240pt}}{We use the abbreviations: m $=$ meters, s $=$ 
seconds, min $=$ minutes, h $=$ hours, and w.r.t. $=$ with respect to.}
\end{tabular}
\label{tanksysinfo}
\end{table}

\subsubsection{Non-linear differentiable continuous model}
The model \eqref{fullequationmodel} is not differentiable, and thus it cannot be linearized about an operating point. Linearization is useful because a controller can be optimized for a linear model more simply than for a non-linear model, and simpler controllers can be adopted more readily in practice. Here, we derive a differentiable continuous approximation for \eqref{fullequationmodel}. We define the model
\begin{subequations}\label{nonlineardiff}
\begin{equation}
x_{t+1} = x_t + \tau f^{\epsilon}(x_t, u_t, w_{t}), \;\;\; t = 0,1,2,\dots,
\end{equation}
where $\epsilon > 0$ is a small positive number, and $f^{\epsilon}$ is given by
\begin{equation}\label{smooth-system}\begin{aligned}
f^{\epsilon}(x,u,w) & \coloneqq \left[ f_{1}^{\epsilon}(x,u,w), f_{2}^{\epsilon}(x,u,w) \right]^{T}\\
f_{1}^{\epsilon}(x,u,w) & \coloneqq  w_\text{r} \cdot a_\text{in} - q_{\text{out}}^{\epsilon}(x_{1}) - q_\text{pump}^{\epsilon}(x,u)\\
f_{2}^{\epsilon}(x,u,w) & \coloneqq  w_\text{r} \cdot a_{2} + q_\text{pump}^{\epsilon}(x,u)  - w_\text{e} - q_\text{drain}^\epsilon(x_2). 
\end{aligned}\end{equation}
\end{subequations}
%
%
The functions $q_{\text{out}}^{\epsilon}$, $q_\text{pump}^{\epsilon}$, and $q_\text{drain}^\epsilon$ are differentiable continuous approximations for $q_{\text{out}}$ \eqref{qout}, $q_\text{pump}$ \eqref{qpump}, and $q_\text{drain}$ \eqref{qdrain}, respectively. Our deviations use sigmoid functions or a smooth square root approximation \cite{duan2016smoothing}. From \cite[Eq. 3, p. 89]{duan2016smoothing}, a smooth square root approximation is defined by $\psi^\epsilon : \mathbb{R} \rightarrow \mathbb{R}$ such that
\begin{equation}
    \psi^\epsilon(y) \coloneqq \begin{cases} \frac{2}{3} \sqrt{\epsilon} & \text{if } y \leq 0 \\ \frac{1}{3\epsilon} y^{3/2} + \frac{2}{3} \sqrt{\epsilon} & \text{if } 0 < y\leq \epsilon\\
\sqrt{y} & \text{if } y > \epsilon. \end{cases} 
\end{equation}
Now, we estimate $q_\text{out}$ \eqref{qout} using $\psi^\epsilon$ as follows:
\begin{equation}\label{smooth-qout}\begin{aligned}
q^{\epsilon}_\text{out}(x_1) & \coloneqq c_{\text{out}} \cdot \psi^\epsilon(\nu(x_1)) \\
\nu(x_1) & \coloneqq \textstyle \frac{x_1}{a_1} - z_\text{o}.
\end{aligned}\end{equation}
The function $q^{\epsilon}_\text{out}$ is continuous and differentiable \cite[p. 89]{duan2016smoothing}. The differentiability can be shown by applying \cite[Thm. 20.10, pp. 160--161]{ross2013elementary} to the definition of the derivative. Intuitively, $q^{\epsilon}_\text{out}$ smooths the ``kink'' that appears in $q_\text{out}$ \eqref{qout} when $\frac{x_1}{a_1} = z_\text{o}$.

\begin{figure}[ht]
\centerline{\includegraphics[width=\columnwidth]{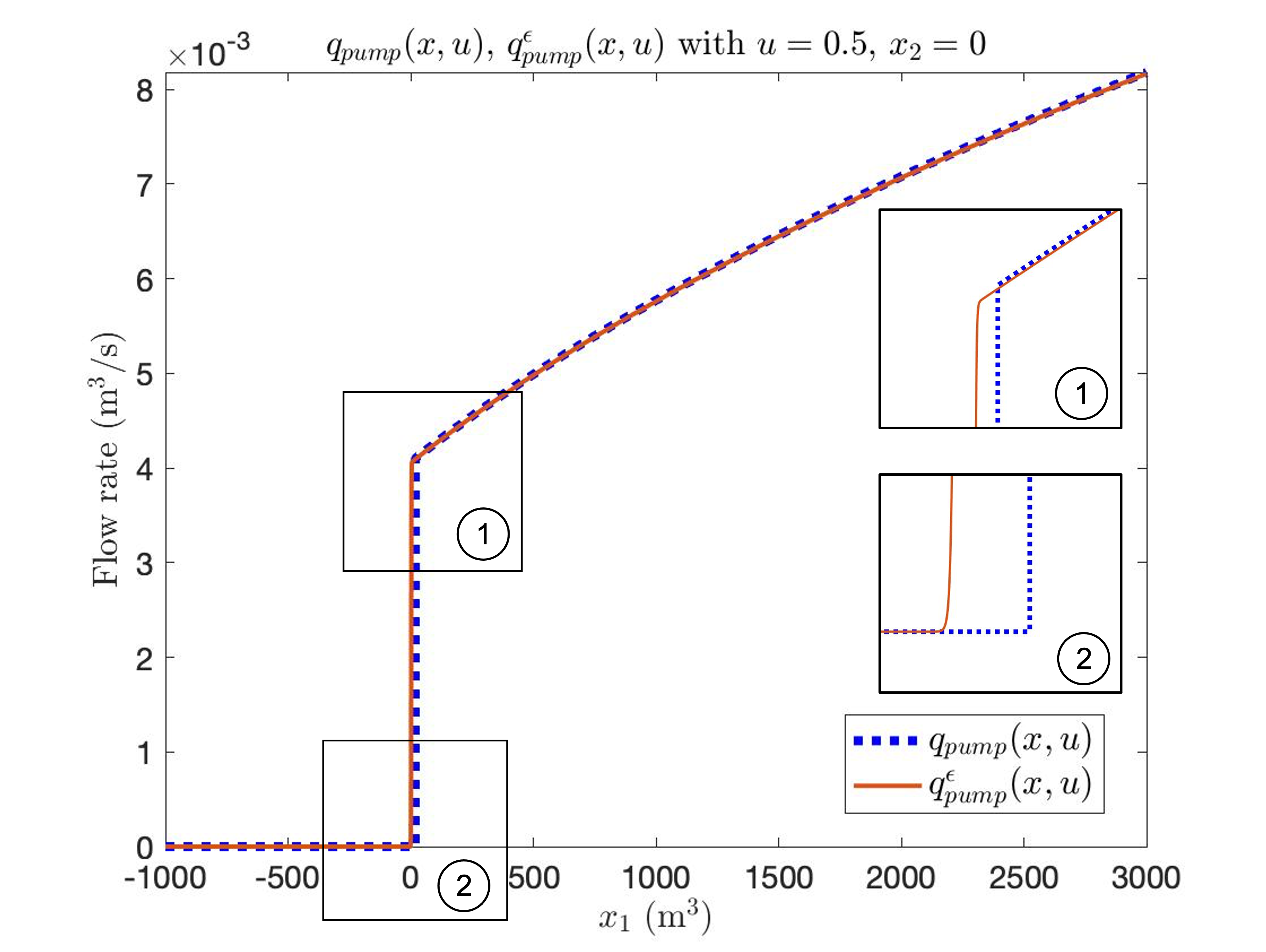}}
\caption{A comparison between $q_{\text{pump}}(x,u)$ \eqref{qpump} and a smooth approximation $q_\text{pump}^\epsilon(x,u)$ \eqref{q_pump_smooth} with $u = 0.5$, $x_2 = 0$, and $\epsilon = 0.5$, as $x_1$ varies.}
\label{compareqpump_figure_2}
\end{figure}

Next, we present our differentiable continuous approximation for $q_\text{pump}(x,u)$ \eqref{qpump}, which is more involved due to the square root together with the additional cases. Our first step is to form a differentiable continuous approximation for the square root in $\bar{q}_{\text{pump},0}(x)$ \eqref{physics-pump-sub} using the approach that we have employed to derive \eqref{smooth-qout},
\begin{subequations}\label{smooth-qpump}
\begin{equation}\label{etaeps}\begin{aligned}
    \eta^\epsilon_{\text{pump}}(x,u) & \coloneqq u \cdot b \cdot  \psi^\epsilon(\rho(x_1) ) \\
    \rho(x_1) & \coloneqq \textstyle \frac{x_1}{a_1} + \hat c - d.
\end{aligned}\end{equation}
Our second step is to model the cases about the sufficiency of water for pumping and the soil moisture using sigmoid functions $\sigma_1^\epsilon$ and $\sigma_2^\epsilon$, respectively.
The function $\sigma_1^\epsilon$ is a differentiable continuous approximation for the case statement about the sufficiency of water for pumping,
\begin{equation}
    \sigma_1^\epsilon(x_1) \hspace{-.5mm} \coloneqq \hspace{-.5mm}  \frac{1}{1\hspace{-.5mm}+\hspace{-.5mm}\exp{(\frac{a_1(z_{\text{pump}} + z_{\text{H}}) - x_1}{\epsilon}})} \hspace{-.5mm}  \approx \hspace{-.5mm}  \begin{cases} 1 & \text{if } \frac{x_{1}}{a_{1}} \hspace{-.5mm} \geq \hspace{-.5mm} z_\text{pump} \hspace{-.5mm} + \hspace{-.5mm}  z_\text{H} \\ 0 & \text{otherwise.} \end{cases}
\end{equation}
Similarly, $\sigma_2^\epsilon$ is a differentiable continuous approximation for the case statement about the soil moisture,
\begin{equation}\label{sigma2}
    \sigma_\text{2}^\epsilon(x_2) \coloneqq \frac{1}{1+\exp{(\frac{x_2-a_2 \cdot z_\text{veg}}{\epsilon}})}  \approx   \begin{cases} 1 & \text{if } \frac{x_{2}}{a_{2}} \hspace{-.5mm} <  z_\text{veg}  \\ 0 & \text{otherwise.} \end{cases}
\end{equation}
We multiply \eqref{etaeps}--\eqref{sigma2} to form a differentiable continuous approximation for $q_\text{pump}(x,u)$ as follows:
\begin{equation}\label{q_pump_smooth}\begin{aligned}
    q_\text{pump}^\epsilon(x,u) & \coloneqq \eta^\epsilon_{\text{pump}}(x,u) \cdot \sigma_1^\epsilon(x_1) \cdot \sigma_2^\epsilon(x_2).
\end{aligned}\end{equation}
\end{subequations}
We compare $q_{\text{pump}}$ \eqref{qpump} and our approximation $q_\text{pump}^\epsilon$ \eqref{q_pump_smooth} numerically in Fig. \ref{compareqpump_figure_2}. 
%

%

Finally, we derive a differentiable continuous approximation for $q_\text{drain}$ \eqref{qdrain} using another sigmoid function to smooth the soil capacity case statement,\vspace{-1mm}
\begin{equation}\label{smooth-qdrain}
    \begin{aligned}
    q_\text{drain}^{\epsilon}(x_2) & \coloneqq K \cdot a_2 \cdot (\textstyle\frac{x_2}{a_2} + z_{\text{soil}})/z_{\text{soil}} \cdot \sigma_3^\epsilon(x_2)\\
    \sigma_3^\epsilon(x_2) & \coloneqq \frac{1}{1 + \exp(\frac{z_{\text{cap}} - x_2}{\epsilon})} \approx \begin{cases} 1 & \text{if }x_2 \geq z_{\text{cap}} \\ 0 & \text{otherwise.} \end{cases}
\end{aligned}\end{equation}
\textcolor{white}{extraspace}
%

\subsubsection{Linear model}\label{linearization_procedure}
We approximate our non-linear differentiable model \eqref{nonlineardiff} near an operating point $p \coloneqq (\bar{x}, \bar{u}, \bar{w}) \in \mathbb{R}^2 \times \mathbb{R} \times \mathbb{R}^2$ at time $t$ by deriving a linear model 
\begin{subequations}\label{linearized}
\begin{equation}
    \tilde x_{i+1} = A_p \tilde x_i + B_p \tilde u_i + C_p \tilde w_i + b_p,
\end{equation}
where $i \in \mathbb{T}_{t}^M \coloneqq \{t, t+1,\dots, t + M-1\}$ and $M \in \mathbb{N}$ is the length of a look-ahead time horizon. The quantities $\tilde x_i \in \mathbb{R}^2$, $\tilde u_i \in \mathbb{R}$, and $\tilde w_i \in \mathbb{R}^2$ are the variations of the state $x_i$, control $u_i$, and disturbance $w_i$ about $p$, respectively,
\begin{equation}
    \tilde x_i  \coloneqq x_i - \bar{x}, \;\; \; \tilde u_i  \coloneqq u_i - \bar{u}, \;\; \; \tilde w_i \coloneqq w_i - \bar{w}.
\end{equation}
We derive the matrices $A_p \in \mathbb{R}^{2 \times 2}$, $B_p \in \mathbb{R}^{2 \times 1}$, $C_p \in \mathbb{R}^{2 \times 1}$, and $b_p \in \mathbb{R}^{2 \times 1}$ by evaluating partial derivatives of $f^\epsilon$ at $p$,
\begin{equation}\begin{aligned}
    A_p & \coloneqq \tau  \frac{\partial f^\epsilon}{\partial x}(p) + I, \;\;\; B_p  \coloneqq \tau  \frac{\partial f^\epsilon}{\partial u}(p), \\
    C_p & \coloneqq \tau  \frac{\partial f^\epsilon}{\partial w}(p), \;\;\; \;\;\;\;\;\;\;\hspace{.5mm} b_p \coloneqq \tau f^\epsilon(p),
\end{aligned}\end{equation}
where $I$ is the $2 \times 2$ identity matrix, $\frac{\partial f^\epsilon}{\partial x}(p)$ is the Jacobian matrix of partial derivatives of $f^\epsilon$ \eqref{nonlineardiff} with respect to $x$ evaluated at $p$, and $\frac{\partial f^\epsilon}{\partial u}(p)$ and $\frac{\partial f^\epsilon}{\partial w}(p)$ are defined similarly.
%
%
\end{subequations}

\subsubsection{Model predictive controller}
Let $t = 0$ and $p = (\bar{x}, \bar{u}, \bar{w})$, where $\bar{x}$ is a given initial state, $\bar{u} = 0$ (no pumping), and $\bar{w} = [0,0]^T$ (no precipitation or evapotranspiration). Let $N \in \mathbb{N}$ be the length of the simulation time horizon. The MPC algorithm proceeds as follows:
\begin{enumerate}
    \item Compute $A_p$, $B_p$, $C_p$, and $b_p$ \eqref{linearized}.
    \item  Compute $\tilde w_{i}$ for each time $i$ in the look-ahead horizon $\mathbb{T}_{t}^M$ using predictions from a local forecast. 
    \item  Compute current and future controls $u_{i}$ for all $i  \in \mathbb{T}_{t}^M$ by minimizing a quadratic cost (to be described) subject to the linear model \eqref{linearized}.  
    \item Apply $u_t$, the control for time $t$ from the previous step, to the non-linear non-differentiable model \eqref{fullequationmodel}.
    \item Proceed to the next time point, i.e., update $t$ by 1.
    \item Measure the current state, and set $\bar{x}$ to this value. Select $\bar{u}$ to be the control from Step 4 and $\bar{w}$ to be the average of the previous disturbances. Using these values, update the operating point $p = (\bar{x},\bar{u},\bar{w})$.
    \item If $t \leq N$, proceed to Step 1; otherwise, stop.
\end{enumerate}

In Step 3, we have chosen a quadratic cost that penalizes the current and future control effort and a deviation between predicted soil moisture and a desired value $z_{\text{veg}}$. Define a vector of future states $X \coloneqq [x_{t+1}^T, \dots, x_{t+M}^T]^T$ and a vector of current and future controls $ U \coloneqq [ u_t,\dots, u_{t+M-1}]^T$. The quadratic cost is a function of $X$ and $U$:
\begin{equation}\begin{aligned}\label{performance_objective_criterion}
    J(X, U) & \coloneqq  c(x_{t+M,2}) + \sum_{i=t}^{t+M-1} c(x_{i,2}) + \lambda  u_{i}^2,
\end{aligned}\end{equation}
where $c(y) \coloneqq \left(\textstyle \frac{y}{a_2} -  z_\text{veg}\right)^2$ with $y \in \mathbb{R}$, the state at time $t$ is $x_t = \bar{x}$, $\lambda > 0$ is a given weight, and $x_{i,2}$ (m\textsuperscript{3}) is the water volume in the green roof at time $i$.
%


\subsection{On/Off Controller}
We consider the following on/off controller. Let $t \in \mathbb{N}$ be the current time, and suppose that $x_t = [x_{t,1},x_{t,2}]^T$ is the state at time $t$. Let $v > 0$ be given. The on/off control $u_t$ equals $v$ if the vegetation requires water and there is sufficient water for pumping but equals zero otherwise,
\begin{equation}
    u_t = \begin{cases} v & \text{if } x_{t,2} < a_2 \cdot z_{\text{veg}} \text{ and $x_{t,1} \geq (z_{\text{pump}} + z_{\text{H}}) a_1$}\\ 0 & \text{otherwise}. \end{cases}
\end{equation}
The on/off controller is considerably simpler to implement, but it does not incorporate information from a forecast.

\section{Outcome}
\begin{figure*}[ht]
\centerline{\includegraphics[width=0.85\textwidth]{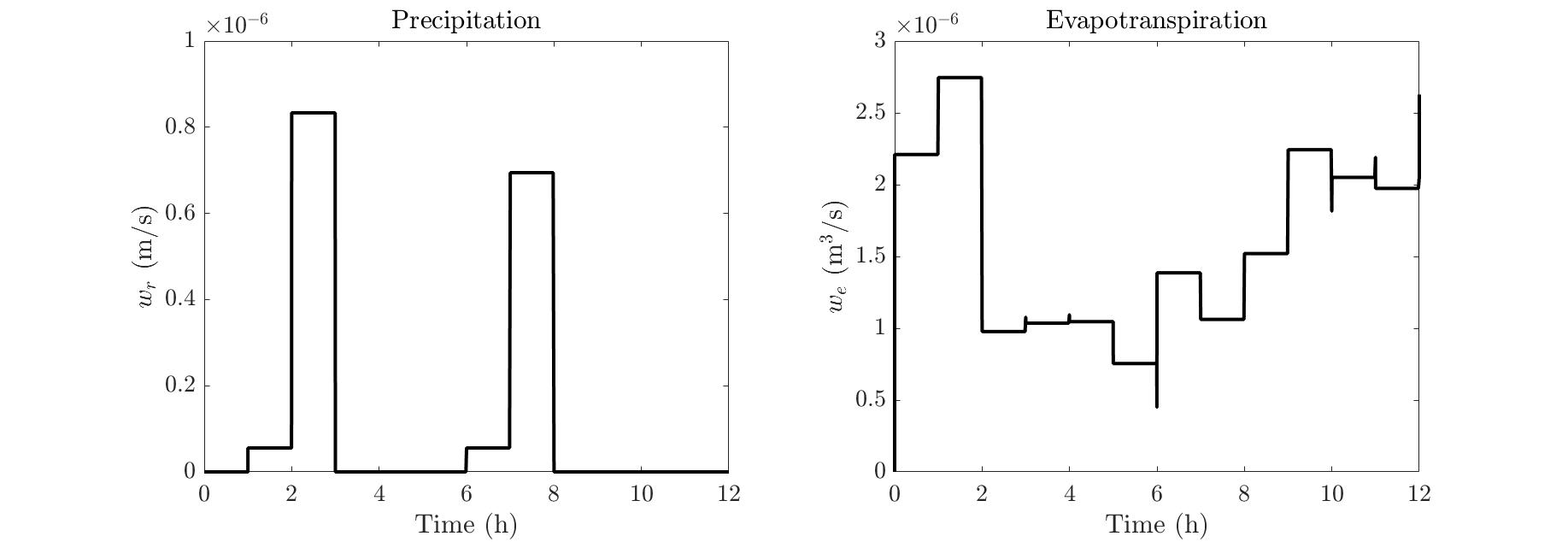}}\vspace{-1mm}
\caption{We present estimated precipitation (left) and evapotranspiration (right) rates from Canadian weather data during a wet month. The precipitation data is from the Toronto City Centre weather station during July 2021 \cite{weatherstation_data}. We have estimated the evapotranspiration rates using the Toronto weather station data \cite{weatherstation_data}, solar irradiance data \cite{solarirradiance_data}, and the Penman-Monteith evapotranspiration method \cite{zotarelli2010step}.}
\label{rainfall_evap}
\end{figure*}

\begin{figure*}[ht]
\centerline{\includegraphics[width=0.85\textwidth]{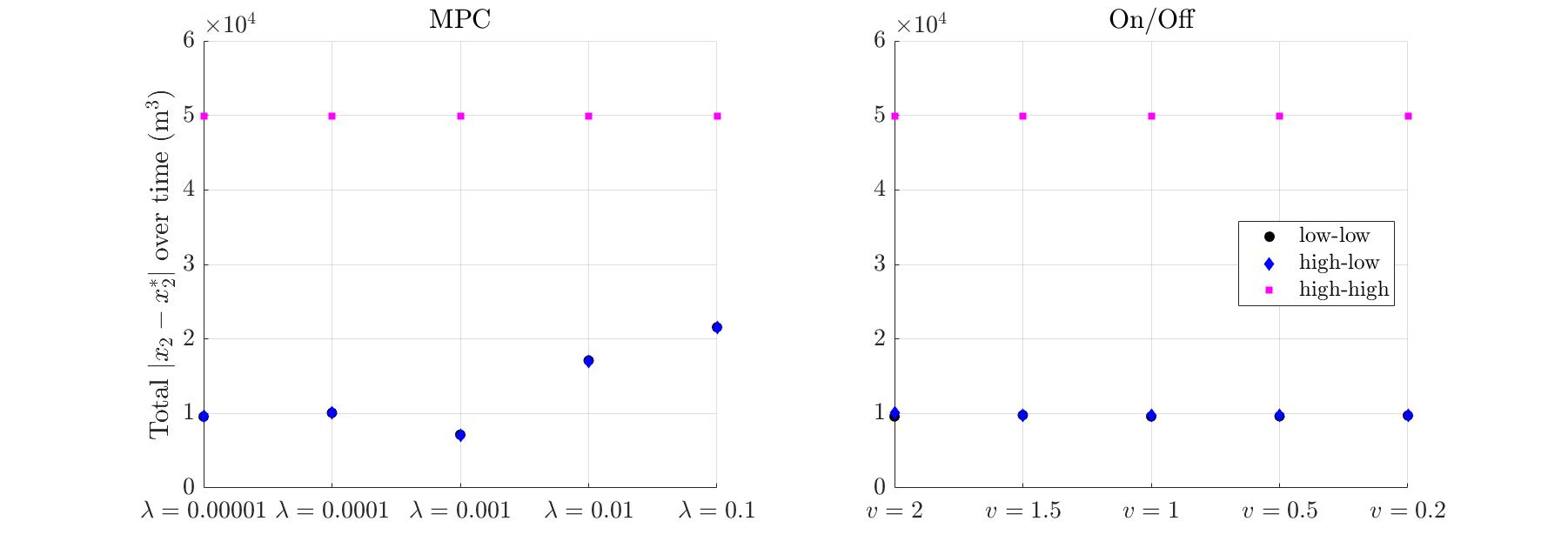}}\vspace{-1mm}
\caption{We present $\sum_{j=0}^N |x_{j,2} - x_2^*|$ under the MPC (left) and on/off controller (right) for each initial state (Table \ref{inittable}). Recall that $x_2^* = a_2 \cdot z_\text{veg}$ is the desired water volume in the green roof to ensure sufficient soil moisture for the vegetation. The points referring to the ``low-low'' and ``high-low'' initial states overlap at the scale shown.}
\label{cum_dev_all_init}
\end{figure*}

To compare the model predictive controller (MPC) and the on/off controller, we have gathered a 12-hour period of time series data from a weather station in Toronto during a wet month (July 2021). The data includes hourly measurements of precipitation, dew point temperature, temperature, and wind speed \cite{weatherstation_data}. We have used the latter three data types and solar irradiance measurements \cite{solarirradiance_data} to estimate an evapotranspiration rate $w_\text{e}$ over time \cite[ET$_0$ equation, p. 2]{zotarelli2010step}.\footnote{The solar irradiance data is from July 2014 in Qu\'{e}bec \cite{solarirradiance_data}; this data type was not available for Toronto in July 2021.} Figure \ref{rainfall_evap} shows the estimated precipitation and evapotranspiration rates. We have evaluated three sets of initial states, where each state is high or low relative to the system's geometry or desired moisture level (Table \ref{inittable}). The term ``low-low'' denotes an initial state with low values for $x_1$ and $x_2$, and the term ``high-low'' denotes an initial state with a high value for $x_1$ and a low value for $x_2$. While the simulations are limited as they assume perfect knowledge of the weather, they offer insights into the anticipated performance of the different controllers. We present the total of $| x_{t,2} - x_2^* |$ over time $t$ for each initial state in Fig. \ref{cum_dev_all_init}, where $x_2^*$ is the desired water volume in the green roof (Table \ref{table-params}). The results show that the MPC with a moderate weight $\lambda$ outperforms the other controllers when pumping is required (Fig. \ref{cum_dev_all_init}). Our code is available from \url{https://github.com/catherineyeh/sustech2021-2}.

\begin{table}[h]
\centering
\caption{Initial states for simulations}
\label{table}
\setlength{\tabcolsep}{3pt}
\begin{tabular}{|p{60pt}|p{80pt}|p{80pt}|}
\hline
\textbf{Name} & \textbf{Initial water volume in the cistern} $x_{0,1}$ (m$^3$) & \textbf{Initial water vol. in the green roof} $x_{0,2}$ (m$^3$)\\  \hline
low-low & $a_1 \cdot z_\text{o} / 1.3$  &  $a_2 \cdot z_\text{veg} / 1.3$ \\ \hline
high-low & $a_1 \cdot z_\text{o} \cdot 1.3$  & $a_2 \cdot z_\text{veg} / 1.3$ \\ \hline
high-high & $a_1 \cdot z_\text{o} \cdot 1.3$  & $a_2 \cdot z_\text{veg} \cdot 1.3$ \\ \hline
\end{tabular}
\label{inittable}
\end{table}
%
%

\begin{figure*}[ht]
\centerline{\includegraphics[width=0.85\textwidth]{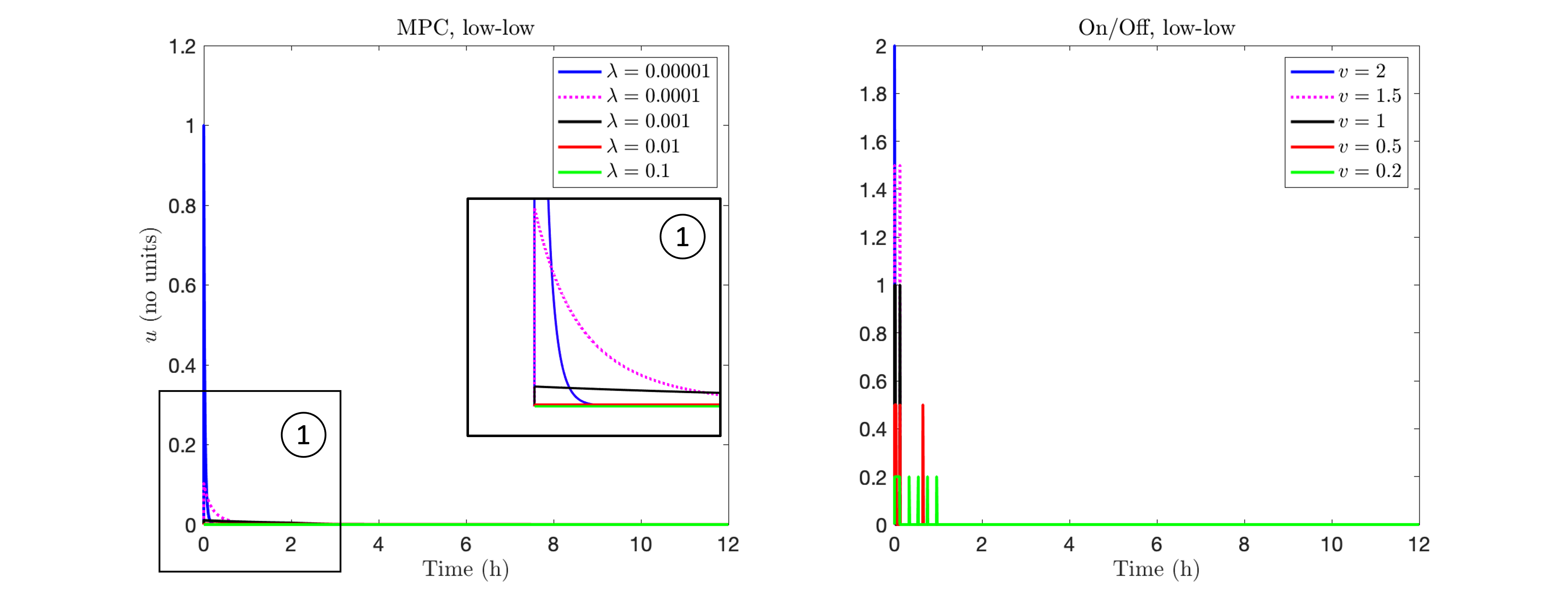}}
\caption{We present the control input $u_t$ versus time $t$ under the MPC (left) and on/off controller (right) for the ``low-low'' initial state (Table \ref{inittable}).}
\label{control_lowlow}
\end{figure*}

\begin{figure*}[ht]
\centerline{\includegraphics[width=0.85\textwidth]{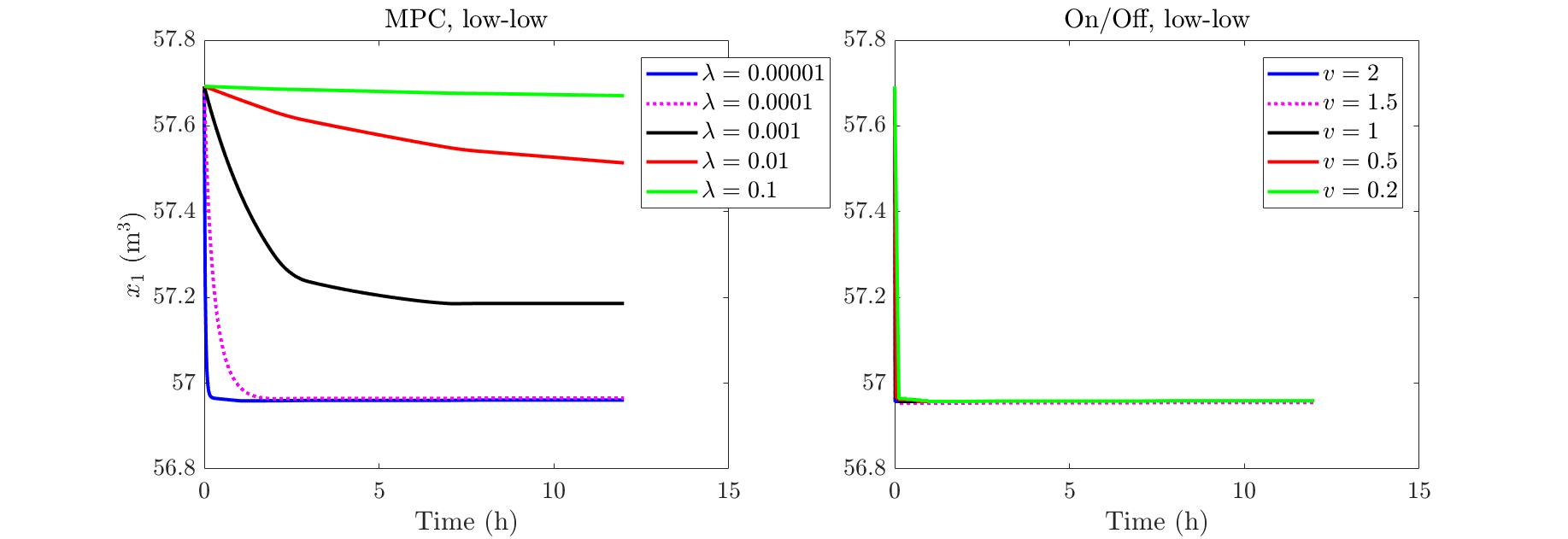}}
\caption{We show the water volume in the cistern $x_{t,1}$ versus time $t$ under the MPC (left) and on/off controller (right) for the ``low-low'' initial state (Table \ref{inittable}).}
\label{state1_lowlow}
\end{figure*}

\begin{figure*}[ht]
\centerline{\includegraphics[width=0.85\textwidth]{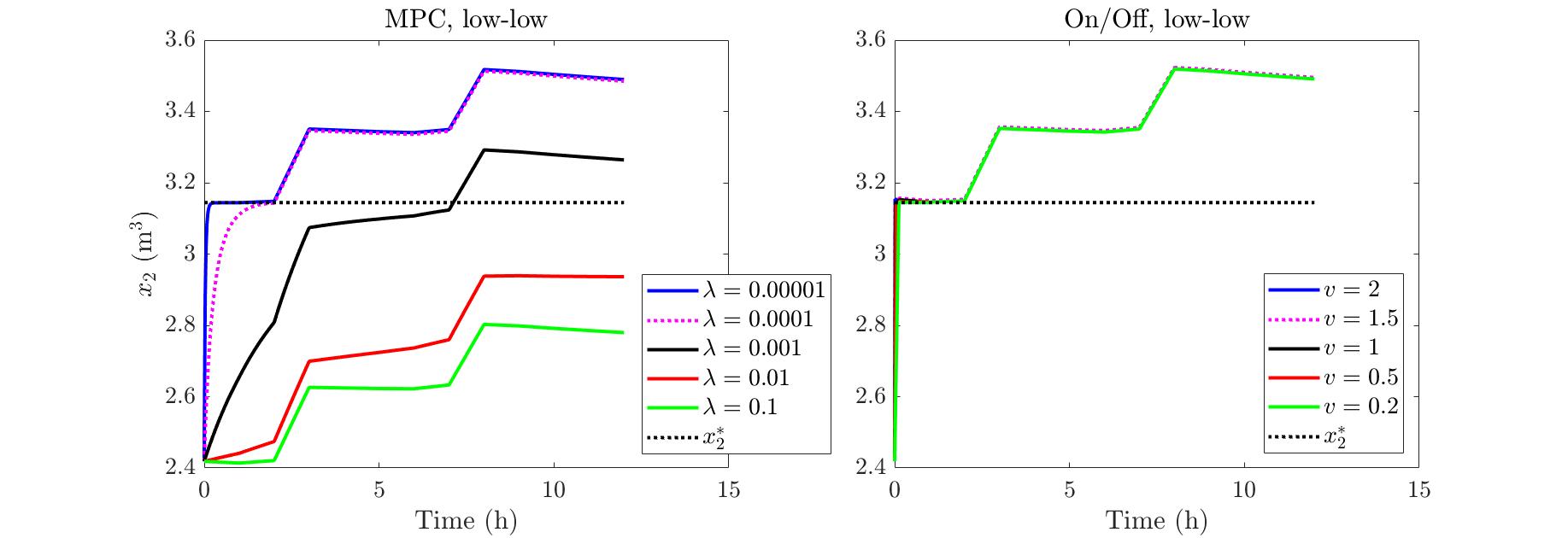}}
\caption{We show the water volume in the green roof $x_{t,2}$ versus time $t$ under the MPC (left) and on/off controller (right) for the ``low-low'' initial state (Table \ref{inittable}). The results for the on/off controller overlap for the different values of $v$.}\vspace{-1mm}
\label{state2_lowlow}
\end{figure*}

First, we discuss the results for the ``low-low'' initial state.
%
%
Both controllers produce high pumping rates early in the time horizon to increase the soil moisture (Fig. \ref{control_lowlow}). However, the MPC permits continuous changes in the pumping rate, whereas the on/off controller inherently lacks this ability (Fig. \ref{control_lowlow}). The behavior of $x_{t,1}$ is shown in Fig. \ref{state1_lowlow}. The behavior of $x_{t,2}$ under the on/off controller for any $v \in \{0.2, 0.5, 1, 1.5, 2\}$ resembles the behavior of $x_{t,2}$ under the MPC when the magnitude of $u$ is penalized the least, i.e., $\lambda = 0.00001$ (Fig. \ref{state2_lowlow}). The on/off controller lacks flexibility and predictive ability, which causes excessive moisture in the green roof after about 2.5 hours (Fig. \ref{state2_lowlow}). The best performance of the MPC occurs for a mid-range penalty, $\lambda = 0.001$, while the performance of the on/off controller is similar for $v$ between 0.2 and 2 (Fig. \ref{cum_dev_all_init}, circles).

\begin{figure*}[ht]
\centerline{\includegraphics[width=0.85\textwidth]{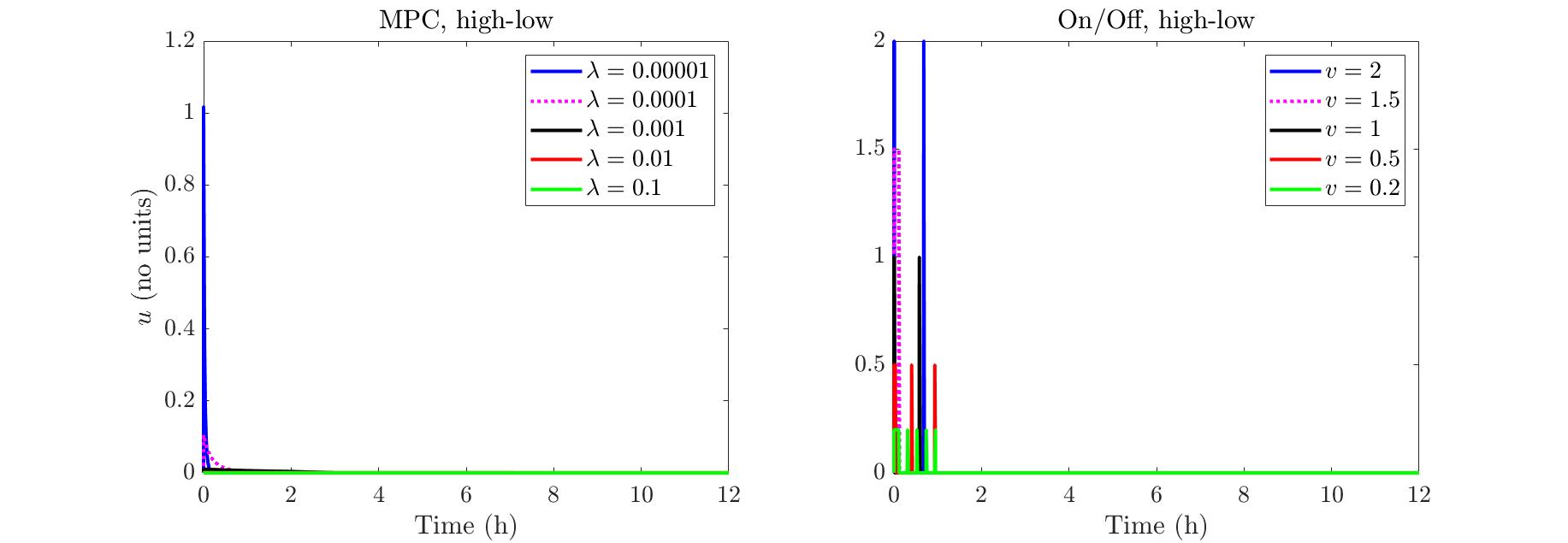}}
\caption{We show the control input $u_t$ versus time $t$ under the MPC (left) and on/off controller (right) for the ``high-low'' initial state (Table \ref{inittable}).}
\label{control_highlow}
\end{figure*}

When the initial water volume in the cistern is high and the initial water volume in the green roof is low (``high-low,'' Table \ref{inittable}), the comparisons of the controllers (Fig. \ref{control_highlow}) and performance (Fig. \ref{cum_dev_all_init}, diamonds) resemble the previous findings  (Fig. \ref{control_lowlow}; Fig. \ref{cum_dev_all_init}, circles). 

When the initial water volumes in both tanks are high (``high-high,'' Table \ref{inittable}), the MPC and on/off controller are zero over time (not shown due to space constraints) because the vegetation does not require irrigation. Thus, in this case, there is no difference in their performance (Fig. \ref{cum_dev_all_init}, squares).

\section{Conclusion}
Using Newtonian physics and smoothing techniques, we have proposed a non-linear differentiable model for reusing stormwater to irrigate a green roof in Toronto. This new model permits linearization and a model predictive controller (MPC) that incorporates weather data. Our simulations indicate that for an appropriate parameter choice, the MPC outperforms an on/off controller, which does not anticipate the weather. 
We see several exciting extensions. 
A stochastic model can strengthen the representation of the weather, and a risk-averse analysis, as in our prior work \cite{chapman2021risk}, can improve sensitivity to severe, random harmful outcomes. 
%
%
We expect that a model predictive controller requires more expensive hardware versus an on/off controller. Future work can include economic assessments and exploring potentially less expensive, hybrid designs, e.g., an on/off controller that anticipates the weather.
Further studies about automatic sensing and control for green and stormwater infrastructure have potential to improve the environmental impact of expanding cities.

\section{Acknowledgement}
The authors gratefully thank Dr. Jennifer Drake and Dr. Darko Joksimovic for many fruitful discussions and for facilitating connections to the GRIT Lab.

\bibliographystyle{IEEEtran}
\bibliography{references_new}

\begin{thebibliography}{10}
\providecommand{\url}[1]{#1}
\csname url@samestyle\endcsname
\providecommand{\newblock}{\relax}
\providecommand{\bibinfo}[2]{#2}
\providecommand{\BIBentrySTDinterwordspacing}{\spaceskip=0pt\relax}
\providecommand{\BIBentryALTinterwordstretchfactor}{4}
\providecommand{\BIBentryALTinterwordspacing}{\spaceskip=\fontdimen2\font plus
\BIBentryALTinterwordstretchfactor\fontdimen3\font minus
  \fontdimen4\font\relax}
\providecommand{\BIBforeignlanguage}[2]{{%
\expandafter\ifx\csname l@#1\endcsname\relax
\typeout{** WARNING: IEEEtran.bst: No hyphenation pattern has been}%
\typeout{** loaded for the language `#1'. Using the pattern for}%
\typeout{** the default language instead.}%
\else
\language=\csname l@#1\endcsname
\fi
#2}}
\providecommand{\BIBdecl}{\relax}
\BIBdecl

\bibitem{parker2019green}
J.~Parker and M.~E. Zingoni~de Baro, ``{Green infrastructure in the urban
  environment: A systematic quantitative review},'' \emph{Sustainability},
  vol.~11, no. 3182, pp. 1--20, 2019.

\bibitem{ERCOLANI2018830}
G.~Ercolani, E.~A. Chiaradia, C.~Gandolfi, F.~Castelli, and D.~Masseroni,
  ``Evaluating performances of green roofs for stormwater runoff mitigation in
  a high flood risk urban catchment,'' \emph{Journal of Hydrology}, vol. 566,
  pp. 830--845, 2018.

\bibitem{BEECHAM2015370}
S.~Beecham and M.~Razzaghmanesh, ``Water quality and quantity investigation of
  green roofs in a dry climate,'' \emph{Water Research}, vol.~70, pp. 370--384,
  2015.

\bibitem{raimondi_becciu_2020}
A.~Raimondi and G.~Becciu, ``Performance of green roofs for rainwater
  control,'' \emph{Water Resources Management}, vol.~35, pp. 99--111, 2020.

\bibitem{greenroofbylaw}
\BIBentryALTinterwordspacing
{City of Toronto}, ``{City of Toronto Green Roof Bylaw},'' 2021, accessed on
  September 12, 2021. [Online]. Available:
  \url{{https://www.toronto.ca/city-government/planning-development/official-plan-guidelines/green-roofs/green-roof-bylaw/}}
\BIBentrySTDinterwordspacing

\bibitem{bylawdoc}
\BIBentryALTinterwordspacing
{Council of the City of Toronto}, ``{Toronto Municipal Code, Chapter 492, Green
  Roofs},'' 2017, accessed on September 12, 2021. [Online]. Available:
  \url{{https://www.toronto.ca/legdocs/municode/1184_492.pdf}}
\BIBentrySTDinterwordspacing

\bibitem{romero2012research}
R.~Romero, J.~L. Muriel, I.~Garc{\'\i}a, and D.~Mu{\~n}oz de~la Pe{\~n}a,
  ``{Research on automatic irrigation control: State of the art and recent
  results},'' \emph{Agricultural Water Management}, vol. 114, pp. 59--66, 2012.

\bibitem{liao2021development}
R.~Liao, S.~Zhang, X.~Zhang, M.~Wang, H.~Wu, and L.~Zhangzhong, ``{Development
  of smart irrigation systems based on real-time soil moisture data in a
  greenhouse: Proof of concept},'' \emph{Agricultural Water Management}, vol.
  245, no. 106632, pp. 1--9, 2021.

\bibitem{chapman2018reachability}
M.~P. Chapman, K.~M. Smith, V.~Cheng, D.~L. Freyberg, and C.~J. Tomlin,
  ``Reachability analysis as a design tool for stormwater systems,'' in
  \emph{Proceedings of the 2018 IEEE Conference on Technologies for
  Sustainability}.\hskip 1em plus 0.5em minus 0.4em\relax IEEE, 2018, pp. 1--8.

\bibitem{sadler2020exploring}
J.~M. Sadler, J.~L. Goodall, M.~Behl, B.~D. Bowes, and M.~M. Morsy, ``Exploring
  real-time control of stormwater systems for mitigating flood risk due to sea
  level rise,'' \emph{Journal of Hydrology}, vol. 583, no. 124571, pp. 1--10,
  2020.

\bibitem{yorkreport}
\BIBentryALTinterwordspacing
{Toronto and Region Conservation Authority}, ``{Evaluation of an Extensive
  Greenroof, York University, Toronto, Ontario},'' July 2006, accessed on
  September 18, 2021. [Online]. Available:
  \url{{https://sustainabletechnologies.ca/app/uploads/2013/03/GR_york_fullreport.pdf}}
\BIBentrySTDinterwordspacing

\bibitem{Aydin2018}
B.~Aydin, S.~Kim, S.~Engineering, and D.~Harp, ``{Designing an Automated
  Sustainable Green Roof System},'' in \emph{Proceedings of the 2018 IISE
  Annual Conference}.\hskip 1em plus 0.5em minus 0.4em\relax Institute of
  Industrial and Systems Engineers, 2018, pp. 251--256.

\bibitem{peng2015seasonal}
L.~L.~H. Peng and C.~Y. Jim, ``{Seasonal and diurnal thermal performance of a
  subtropical extensive green roof: The impacts of background weather
  parameters},'' \emph{Sustainability}, vol.~7, pp. 11\,098--11\,113, 2015.

\bibitem{TSANG2016360}
S.~Tsang and C.~Jim, ``Applying artificial intelligence modeling to optimize
  green roof irrigation,'' \emph{Energy and Buildings}, vol. 127, pp. 360--369,
  2016.

\bibitem{torontocso}
\BIBentryALTinterwordspacing
{City of Toronto}, ``{Combined Sewer Overflows},'' 2021, accessed on October
  20, 2021. [Online]. Available:
  \url{https://www.toronto.ca/services-payments/water-environment/managing-rain-melted-snow/what-is-stormwater-where-does-it-go/combined-sewer-overflows/}
\BIBentrySTDinterwordspacing

\bibitem{duan2016smoothing}
Y.~Duan and S.~Lian, ``{Smoothing approximation to the square-root exact
  penalty function},'' \emph{Journal of Systems Science and Information},
  vol.~4, no.~1, pp. 87--96, 2016.

\bibitem{zotarelli2010step}
\BIBentryALTinterwordspacing
L.~Zotarelli, M.~D. Dukes, C.~C. Romero, K.~W. Migliaccio, and K.~T. Morgan,
  ``{Step by step calculation of the Penman-Monteith Evapotranspiration (FAO-56
  Method)},'' Institute of Food and Agricultural Sciences, University of
  Florida, 2010. [Online]. Available:
  \url{https://edis.ifas.ufl.edu/pdf/AE/AE45900.pdf}
\BIBentrySTDinterwordspacing

\bibitem{white2011fluid}
F.~M. White, \emph{Fluid Mechanics}, 7th~ed., ser. {McGraw-Hill Series in
  Mechanical Engineering}.\hskip 1em plus 0.5em minus 0.4em\relax New York, NY:
  McGraw-Hill, 2011.

\bibitem{selker_or_2018}
\BIBentryALTinterwordspacing
J.~Selker and D.~Or, \emph{Soil Hydrology and Biophysics}.\hskip 1em plus 0.5em
  minus 0.4em\relax Corvallis, OR: Oregon State University, 2018. [Online].
  Available:
  \url{https://open.oregonstate.education/soilhydrologyandbiophysics/}
\BIBentrySTDinterwordspacing

\bibitem{al-kharabsheh_azzam_2019}
N.~Al-Kharabsheh and R.~Azzam, ``Mass transport of nitrate in soil by utilizing
  the optimized diffusion cell and emission-transmission-immission concept,''
  \emph{Polish Journal of Environmental Studies}, vol.~28, no.~4, pp.
  2553--2563, 2019.

\bibitem{melvin_yonts_2009}
\BIBentryALTinterwordspacing
S.~R. Melvin and C.~D. Yonts, ``{Irrigation Scheduling: Checkbook Method},''
  University of Nebraska--Lincoln Extension, Division of the Institute of
  Agriculture and Natural Resources, 2009. [Online]. Available:
  \url{https://extensionpublications.unl.edu/assets/pdf/ec709.pdf}
\BIBentrySTDinterwordspacing

\bibitem{ross2013elementary}
K.~A. Ross, \emph{Elementary Analysis}, 2nd~ed.\hskip 1em plus 0.5em minus
  0.4em\relax New York, NY: Springer, 2013.

\bibitem{weatherstation_data}
\BIBentryALTinterwordspacing
{Government of Canada}, ``{Historical Data},'' {Toronto City Centre weather
  station data in July 2021, website accessed in October 2021}. [Online].
  Available:
  \url{https://climate.weather.gc.ca/historical_data/search_historic_data_e.html}
\BIBentrySTDinterwordspacing

\bibitem{solarirradiance_data}
\BIBentryALTinterwordspacing
------, ``{High-Resolution Solar Radiation Datasets},'' Feb. 2020, location of
  sensors: Varennes, Qu\'{e}bec; measurements are from July 2014. [Online].
  Available:
  \url{https://www.nrcan.gc.ca/energy/renewable-electricity/solar-photovoltaic/18409}
\BIBentrySTDinterwordspacing

\bibitem{chapman2021risk}
M.~P. Chapman, R.~Bonalli, K.~M. Smith, I.~Yang, M.~Pavone, and C.~J. Tomlin,
  ``{Risk-sensitive safety analysis using Conditional Value-at-Risk},'' 2021,
  conditionally accepted by \emph{IEEE Transactions on Automatic Control}.

\end{thebibliography}

\end{document}